# Nanoscale grains, high irreversibility field, and large critical current density as a function of high energy ball milling time in C-doped magnesium diboride

B J Senkowicz[1], R J Mungall, Y Zhu[2], J Jiang[1], P M Voyles[2], E E Hellstrom[1,2], and D C Larbalestier[1]

[1] *Applied Superconductivity Center at National High Magnetic Field Laboratory, Florida State University, Tallahassee, FL 32310*

[2] *Department of Materials Science and Engineering, University of Wisconsin – Madison, Madison, WI 53706*

Magnesium diboride ($MgB_2$) powder was mechanically alloyed by high energy ball milling with C to a composition of $Mg(B_{0.95}C_{0.05})_2$ and then sintered at 1000°C in a hot isostatic press. Milling times varied from 1 minute to 3000 minutes. Full C incorporation required only 30-60 min of milling. Grain size of sintered samples decreased with increased milling time to <30 nm for 20-50 hrs of milling. Milling had a weak detrimental effect on connectivity. Strong irreversibility field (H*) increase (from 13.3 T to 17.2 T at 4.2 K) due to increased milling time was observed and correlated linearly with inverse grain size (1/d). As a result, high field $J_c$ benefited greatly from lengthy powder milling. $J_c$(8 T, 4.2 K) peaked at > 80,000 A/cm$^2$ with 1200 min of milling compared with only ~ 26,000 A/cm$^2$ for 60 min of milling. This non-compositional performance increase is attributed to grain refinement of the unsintered powder by milling, and to the probable suppression of grain growth by milling-induced MgO nano-dispersions.

## Introduction

The literature on $MgB_2$ contains numerous reports of doping with a variety of elements; most notably C and C-containing species, which can improve $H_{c2}$.[1-13] The vast majority of $MgB_2$ doping studies are carried out with the dopant present during the initial Mg + B → $MgB_2$ reaction. That method is best suited for *in situ* conductor fabrication. In contrast, despite the success of *ex situ* conductors [14,15], comparatively little work has been published regarding doping of pre-reacted $MgB_2$.

Our previous work [11,12] showed that C can be successfully incorporated into the pre-reacted $MgB_2$ lattice in concentrations up to X~0.06 (in $Mg(B_{1-X}C_X)_2$) by high energy ball milling for 10 hours followed by heat treatment at 1000°C. This procedure enhanced $H_{c2}$ similarly to other bulk and single crystal C-doping methods [16-21] and resulted in H*(4.2 K) > 16 T, and $J_c$(8 T, 4.2 K) > 50,000 A/cm$^2$. The high $J_c$ was partly due to non-compositional factors associated with milling, particularly the very *small grain size* (<50 nm). Since those samples all had identical milling time ($t_{mill}$=10 hrs) the timescale required for C-incorporation and grain refinement remained unclear. One of the key observations in that work was higher than expected H*(4.2 K) in very lightly C-doped (nominally pure) samples.

As part of a recent review article, Braccini et al. [14] reported on the effect of high-energy ball milling on $J_c$ of undoped *ex situ* tapes. They found that increased (but unspecified) milling time had only a small effect on low-field $J_c$ of the heat treated samples, but that the irreversibility field (H*) increased from 5.7 T to 11 T with H⊥tape surface and 8.8 to 12 T with H$^{//}$tape surface. This was due to "lowered… average grain size", but the grain size was not quantified. They suggested that grain refinement by milling results in less rolling-induced texture anisotropy in their tapes, and that fine grains may beneficially change current percolation paths.

Non-compositional effects on H* have also been observed in *in situ* samples. Herrmann et al. [4] recently reported H*(4.2 K) > 15 T for nominally undoped samples (and even higher for C-doped) as a result of extremely fine grain size obtained by ball milling for 50 hours. Since grain boundary flux pinning does not account for such a result, and in view of increased $H_{c2}$ and decreased $T_c$, they concluded that grain boundary electron scattering drove the material into the dirty limit regime.

Kiuchi et al.[22] and Yamamoto et al. [23, 24] observed H* increases in undoped, *in-situ* samples as a result of low reaction temperature. Kiuchi et al. attributed the effect to "small grains" but did not attempt to estimate grain size. They also pointed out that $H_{c2}$ is much more important than either the flux pinning force or connectivity in determining $J_c$ at high magnetic fields. We will also make that point. Yamamoto et al. discovered a link between H* and the broadness (FWHM) of the (110) reflection in $MgB_2$ x-ray diffraction patterns, indicating that some combination of disorder, strain, and/or grain size strongly affected H*. They suggested "that distortion of [the] honeycomb boron sheet is directly associated with the intraband scattering resulting in an enhancement of grain boundary flux pinning."[24] We will suggest that the primary significance of the x-ray diffraction peak broadening in our study is to indicate fine grain size, and that it is the very high grain boundary density which increases H* and decreases $H_{c2}$ anisotropy

Grain refinement is therefore known to improve performance not just by increased flux pinning, but by increasing H* in some way probably related to increased $H_{c2}$. This work elucidates those effects by investigating the evolution of composition, grain size, and performance in sintered *ex-situ* samples as a function of ball milling time.

We will show that full dopant incorporation was achieved at short milling time ($t_{mill}$) = 60 min, yet further milling substantially improved superconducting properties. Extended milling times (600 min or more) resulted in 20-30 nm grains after heat treatment, which correlated directly to increased irreversibility field (H*). The result was substantially enhanced performance for milling times up to 20 hours.

**Experimental Procedure**

High energy ball milling was conducted in a SPEX 8000M mill with WC jar and media, mixing pre-reacted $MgB_2$ from Alfa-Aesar with powdered graphite and a stoichiometric amount of Mg to make the nominal bulk composition $Mg(B_{0.96}C_{0.04})_2$. A small amount of C was present in the as-received powder, so that the overall bulk composition was approximately $Mg(B_{0.95}C_{0.05})_2$. Powder milling times were 1, 15, 60, 300, 600, 1200, and 3000 min. No sample was made for $t_{mill}$ = 0 min because it was assumed that the powders would not be homogeneously mixed.

After milling, powders were pressed into pellets, sealed in evacuated stainless steel tubes, and hot isostatic pressed (HIP) at a maximum T and P of $1000^oC$ / ~30kpsi for 200 min. All pre-heat treatment work was carried out in a nitrogen-filled glove box equipped with active oxygen and water scavenging. After HIP heat treatment, hard (> 1000 $H_V$) and (~90%) dense samples were sectioned with a diamond saw [12].

Resistive $H_{c2}$ measurements were made in magnetic fields up to 9 T in a Quantum Design Physical Properties Measurement System (PPMS) and high-field measurements were made in a 33 T magnet at the National High Magnetic Field Laboratory (NHMFL) in Tallahassee, FL. $H_{c2}$ was extracted from swept field (0.6 T/min) 4-point resistive transitions using a measurement current of 5 mA in the PPMS, and 10 mA in the high field magnet. Cross sectional area was ~1 $mm^2$ with gage length ~2 mm and R ~ 1 mΩ at 40 K, making the measurement current density ~0.5 - 1 $A/cm^2$. Because small current measurements report zero resistance while a sufficiently large percolative superconducting path exists, the $H_{c2}$ values thus measured represent the most favorably oriented crystals, which are those with ab-planes parallel to the field direction, referred to here as $H_{c2}^{//}$.

$J_c$ values were derived from M-H hysteresis loops measured in a vibrating sample magnetometer (VSM) applying the Bean model according to the expression

$$(1) \qquad M = \frac{3b-d}{12b} J_c d$$

where M is the volumetric magnetization, and *b* and *d* are the sample dimensions perpendicular to the field direction, with $b > d$.[25,26] Typical values of *b* and *d* were 1 mm and 0.5 mm, which at a magnetic field ramp rate of 0.6 T/min, defines the electric field at ~ 0.1 μV/cm. These measurements were also used to determine the irreversibility field (H*). $T_c$ was determined by m(T) measurements in a SQUID magnetometer using an applied field of 5 mT.

X-ray diffraction was performed with both an area-detector equipped system to identify phases and determine texture, and a high-resolution instrument for precise determination of peak positions, both using Cu Kα radiation. In these measurements peak positions and FWHM were determined by least squares data fit using FullProf software. The Nelson-Riley method was used to determine accurate lattice parameters, and the Williamson-Hall analysis was employed in grain size calculations. Analytical transmission electron microscopy (TEM) was performed on a Phillips CM200UT TEM.

**Results**

X-Ray diffraction (Fig 1) after heat treatment showed randomly oriented $MgB_2$. As milling time increased, the $MgB_2$ peaks became weaker and broader. Small amounts of MgO were identified, particularly in the samples milled for short times of 60 min or less. In samples milled longer than about 15 min, WC, WB, and $MgB_2C_2$ were evident. The WC signature increased in prominence with increased milling time. High resolution XRD was carried out to determine precise peak positions and widths. The Nelson-Riley method was applied to the peak positions to calculate a-axis lattice parameter values (Table I) that were then compared to the C-doped single crystal data of Kazakov et al.[19] to estimate the actual (rather than nominal) C-content of the lattice, as described in the discussion section. Table I gives FWHM values for the (100) and (002) reflections, and calculated lattice parameters from the high-resolution patterns. Rapid a-parameter decrease with $t_{mill}$ was observed for the first hour of milling, with $a$ decreasing from ~3.076 Å to ~3.0695 Å. For $t_{mill} \geq 60$ min, the a-axis lattice parameter was nearly constant, indicating that the lattice C composition was also constant, despite the presence of more WC in the system. Peak broadening mostly occurred within the first 300 – 600 minutes of milling, with FWHM increasing from ~0.25° to ~0.4° in 2θ for the (100) and (002) reflections. Later we will discuss using the Williamson-Hall analysis to estimate grain size and microstrain.

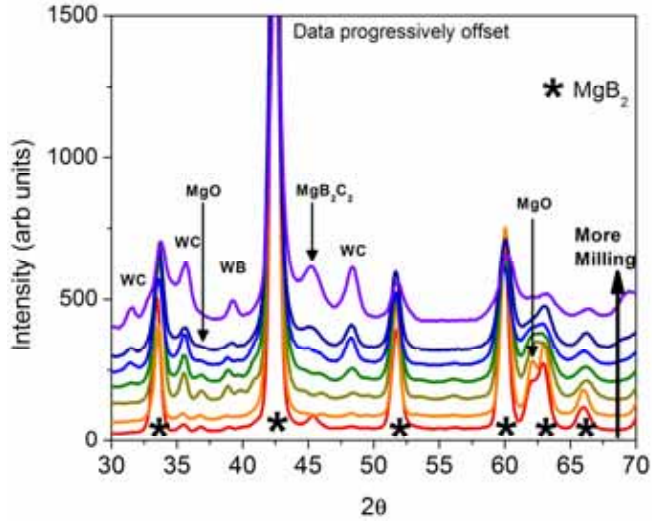

**Figure 1** – Low-resolution x-ray diffraction patterns for each sample after sintering, progressively offset upwards on the Y-axis with short $t_{mill}$ at bottom.

**Table I** – High resolution x-ray diffraction results from sintered samples.

| Milling Time (min) | a-parameter (Å) | Estimated Uncertainty (+-) (Å) | FWHM (100) (Degrees 2θ) | FWHM (002) (Degrees 2θ) |
|---|---|---|---|---|
| 1 | 3.0758 | 0.0040 | 0.25 | 0.26 |
| 15 | 3.0746 | 0.0031 | 0.27 | 0.31 |
| 60 | 3.0695 | 0.0036 | 0.31 | 0.33 |
| 300 | 3.0698 | 0.0061 | 0.38 | 0.44 |
| 600 | 3.0683 | 0.0032 | 0.45 | 0.52 |
| 1200 | 3.0685 | 0.0060 | 0.37 | 0.42 |
| 3000 | 3.0685 | 0.0112 | 0.48 | 0.52 |

Transmission electron microscopy (Fig. 2) was done on the 1200 min (Fig. 2a) and 3000 min (Fig. 2b,c) samples. It revealed an average grain size of 20 to 30 nm in the 1200 min sample. The 3000 min sample appeared to have even smaller grains, but some much larger (0.5 – 1 μm), comparatively strain-free grains were also observed. Larger amounts of WC rubble were observed in the 3000 min sample (6.1 % projected area, Fig 2c) than the 1200 min sample. Trace amounts of cobalt from the milling media were detected by energy dispersive spectroscopy (EDS) in fine grained regions of both samples, but no Co was found within large, single-phase $MgB_2$ areas.

Numerous potential obstacles to current flow were observed in the 3000 min sample, (Figure 2c) including WC which appears as dark regions, and porosity, amorphous regions, and other second phases which appear as bright regions. Electron diffraction patterns on earlier samples [12] indicated that MgO is the most prevalent phase after $MgB_2$ and WC. However, second phases did *not* form continuous layers at grain boundaries, as occasionally observed in other work. [27]

Zero-field-cooled, warming magnetization curves (5 mT field) are shown in Fig. 3. $T_c$ was defined at -m = -0.02 m(5 K). $T_c$ decreased with increased milling time, but the relationship was not simple. A rapid $T_c$ decrease from 37.1 to 32.6 K was observed up through the 300 min sample, but $T_c$ decreased only 1.2 K from 32.6 K to 31.4 K between $t_{mill}$ = 300 and 1200 min. The 3000 min sample had $T_c$ = 24.6 K as defined above, but a small fraction of the sample (~1.5% of signal) had higher $T_c$ ~ 31 K (Fig 3b).

All samples exhibited metallic-type resistivity behavior. Resistive properties are shown in Table II along with SQUID-derived (not resistive) $T_c$ from Fig 3. Measured $\rho(40\ K)$ and $\rho(300\ K)$ both increased with increasing milling time. At 40 K, $\rho(1\ min) = 14\ \mu\Omega$-cm and $\rho(3000\ \text{min}) = 281\ \mu\Omega\text{-cm}$. RRR decreased from 2.36 to 1.15 as milling time increased over the same range. $\Delta\rho = \rho(300\ K) - \rho(40\ K)$ increased with longer milling time, but only by a factor of ~2 over the entire sample set. For short $t_{mill}$ samples (up to 15 min) the values of $\rho(40\ K) = 16$ $\mu\Omega$-cm, RRR = 2.13, and $T_c$ = 36.8 K are quite reasonable for C-doped bulks. However, the trends of increasing $\rho$ and decreasing $T_c$ continue through the sample set long after the lattice parameters cease to change. We will show that as long as $\Delta\rho$ remained fairly constant (up to $t_{mill}$ = 1200 min) increasing resistivity correlates with *increased* $J_c$(8 T, 4.2 K).

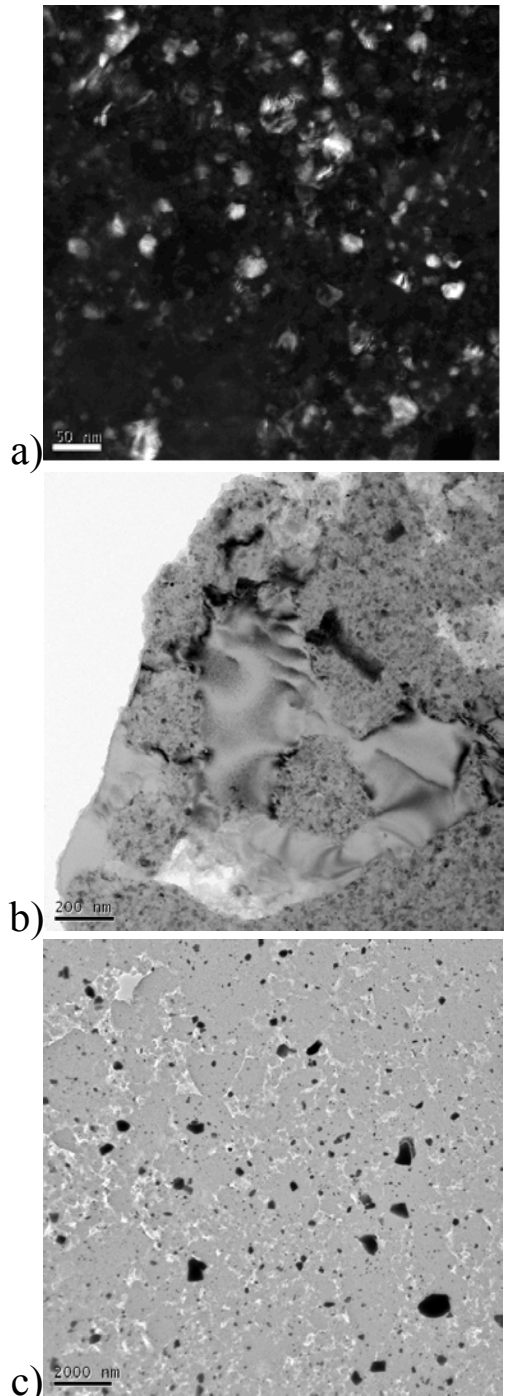

**Figure 2** – (a) Dark-field and (b,c) bright-field TEM images showing grain related diffraction contrast in sintered samples made from powder milled for a) 1200 min and b and c) 3000 min. Scale bar size is (a) 50 nm, (b) 200 nm and (c) 2000 nm. The average grain size for the 1200 min sample was <30 nm. Typical grain size for the 3000 min sample was smaller, but large grains also exist. Dark areas in (c) are WC and constitute 6.1% by area. Bright regions in (c) occupy 14% by area and are a combination of porosity, amorphous or very fine grained regions, and second phases.

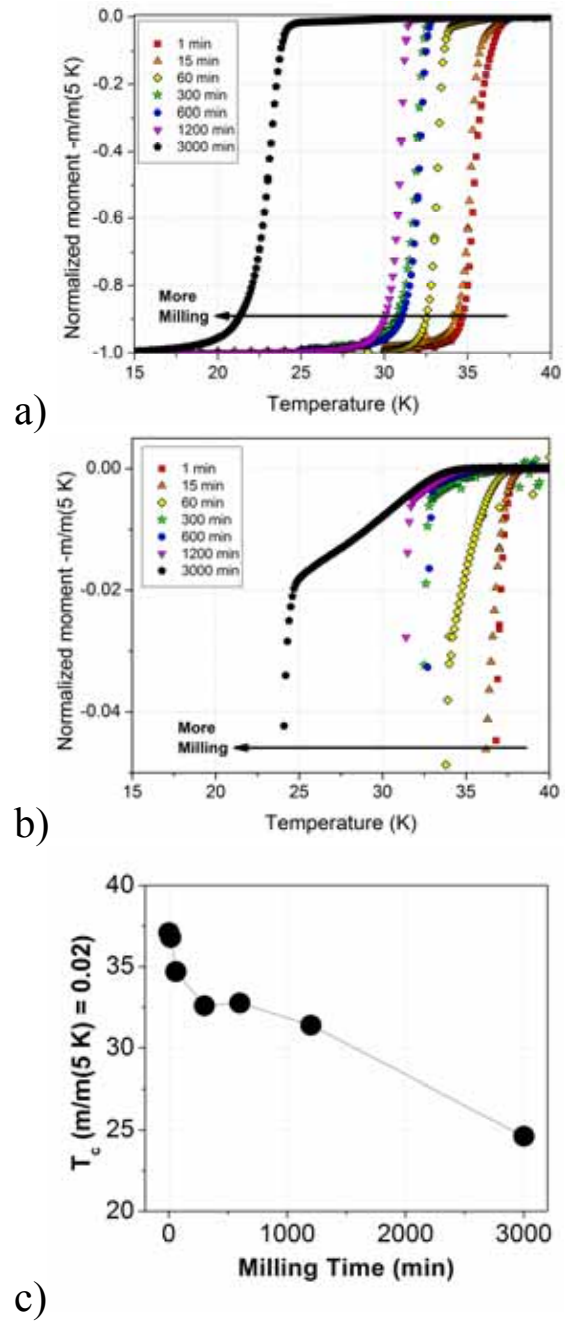

**Figure 3** – a) and b) $T_c$ traces at 5 mT applied field. The complete transitions are shown in a) while only the upper portion of the transitions is shown in b). C) $T_c$ at 2% below onset ($-m/m(5\ K) = -0.02$) as a function of milling time.

**Table II** – Resistive properties and $T_c$ as a function of milling time.

| Milling Time (minutes) | $\rho(300\ K)$ ($\mu\Omega$-cm) | $\rho(40\ K)$ ($\mu\Omega$-cm) | RRR | $\Delta\rho$ ($\mu\Omega$-cm) | $T_c$ (K) |
|---|---|---|---|---|---|
| 1 | 33 | 14 | 2.36 | 19 | 37.1 |
| 15 | 34 | 16 | 2.13 | 18 | 36.8 |
| 60 | 56 | 33 | 1.70 | 23 | 34.7 |
| 300 | 82 | 57 | 1.44 | 25 | 32.6 |
| 600 | 97 | 72 | 1.35 | 25 | 32.8 |
| 1200 | 132 | 104 | 1.27 | 28 | 31.4 |
| 3000 | 322 | 281 | 1.15 | 41 | 24.6 |

Figure 4 shows $H_{c2}$ extracted from resistive transitions taken at 90% of $\rho(40\ K)$. This corresponds to $H_{c2}$ for those grains with H parallel to the Mg and B planes, here designated $H_{c2}^{//}$ as opposed to $H_{c2}^{\perp}$, associated with grains having H parallel to the c-axis and perpendicular to the ab planes. Progressive $T_c$ suppression by milling

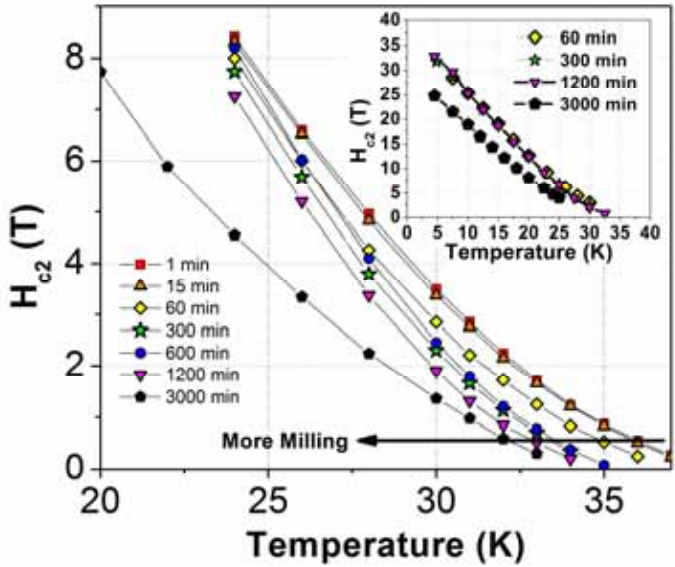

**Figure 4** – $H_{c2}^{//}(T)$ deduced from resistivity curves. Inset includes high-field data for 60, 300, 1200 and 3000 min samples. 60, 300, and 1200 min high-field data fall on the same line and are difficult to distinguish.

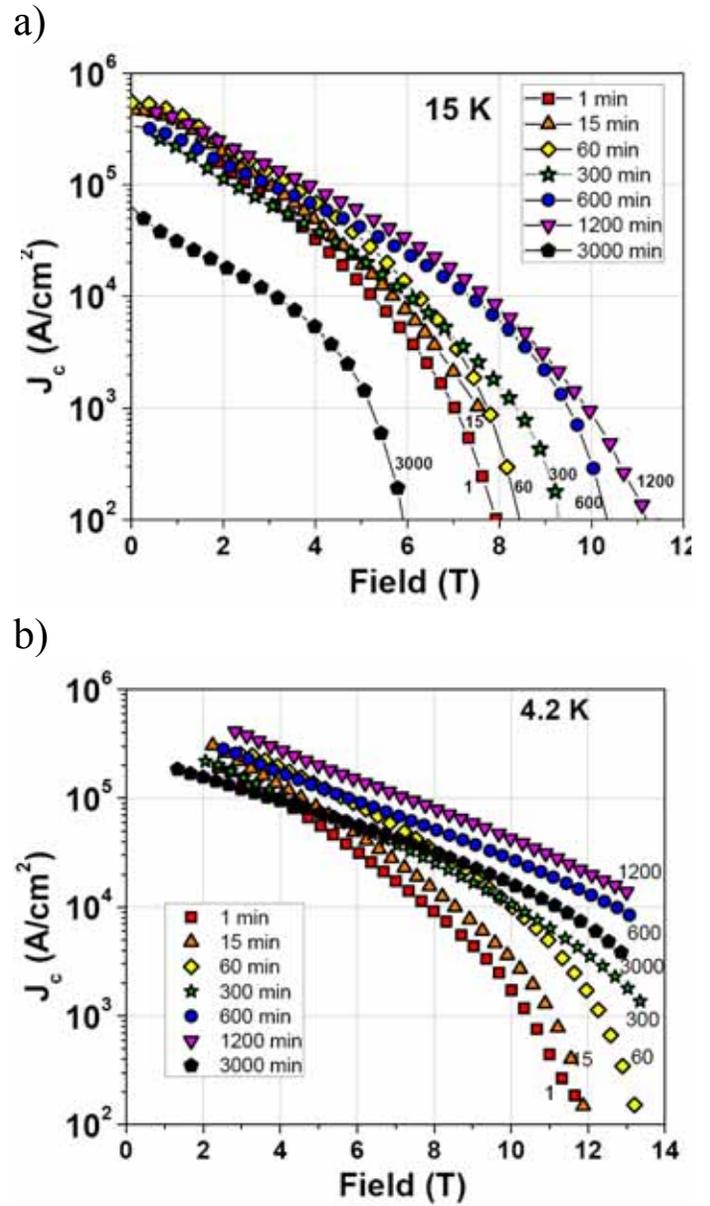

**Figure 5** – $J_c$(H) at a) 15 K and b) 4.2 K.

(see Table II) is also evident in the $H_{c2}$ plots. $T_c$ values taken from Fig 4 are higher than those in Fig 3c because the resistive measurement represents only the highest $T_c$ crystals. These resistive $T_c$ data are therefore approximately equivalent to the first onset of superconductivity in Fig 3b. High-field measurements on the 60, 300, and 1200 min samples (indistinguishable in Fig. 4 inset) found nearly identical $H_{c2}(T)$ with $H_{c2}$(4.2 K) = 33 T and $H_{c2}(0)$ judged by linear extrapolations was ~38 T. The 3000 min sample had both the lowest $T_c$ and the most gradual $H_{c2}(T)$ curve slope, with $H_{c2}(t_{mill}$=3000 min, 4.5 K) = 25 T, as also shown in Fig. 4 inset. At 4.5 K, the transition breadths (H(90%) – H(10%)) were ~6 T, 5.6 T, 6 T, and 4.4 T for $t_{mill}$ = 60, 300, 1200, and 3000 min, respectively. Transition sharpening due to heavy milling may be related to anisotropy reduction, discussed below.

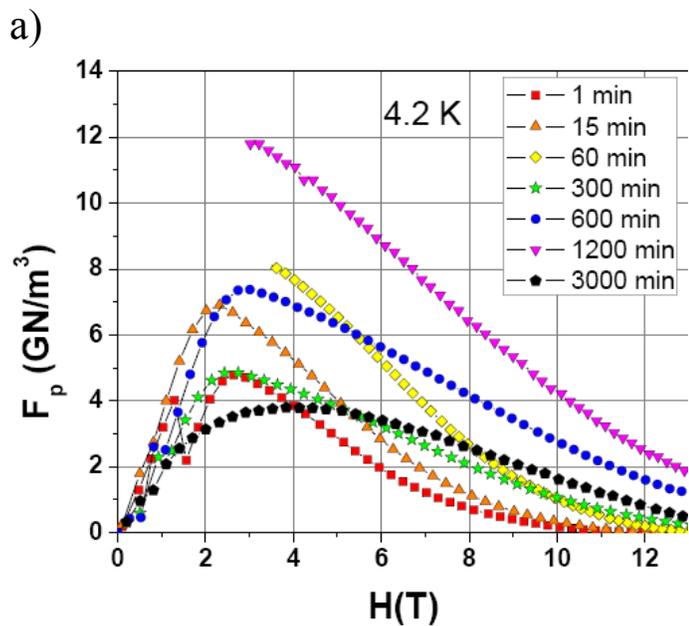

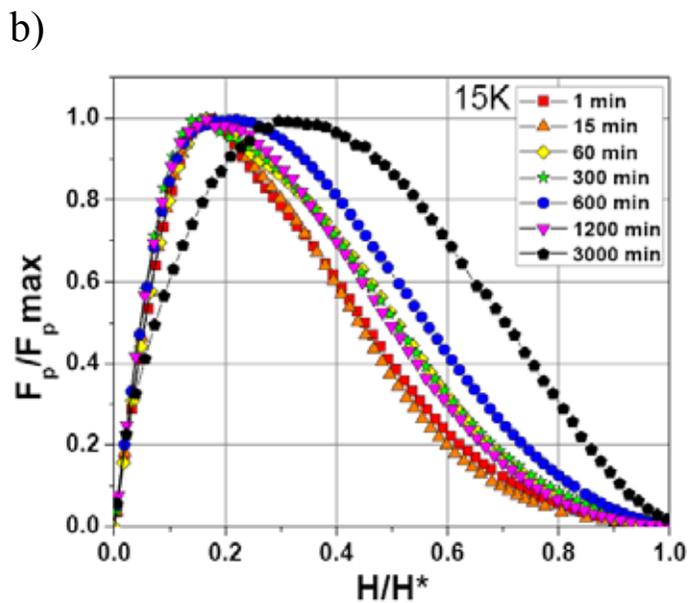

**Figure 6** – Pinning Force Curves a) 4.2 K and b) 15 K. Curves in b) have been normalized to maximum $F_p$ and irreversibility field ($H^*$).

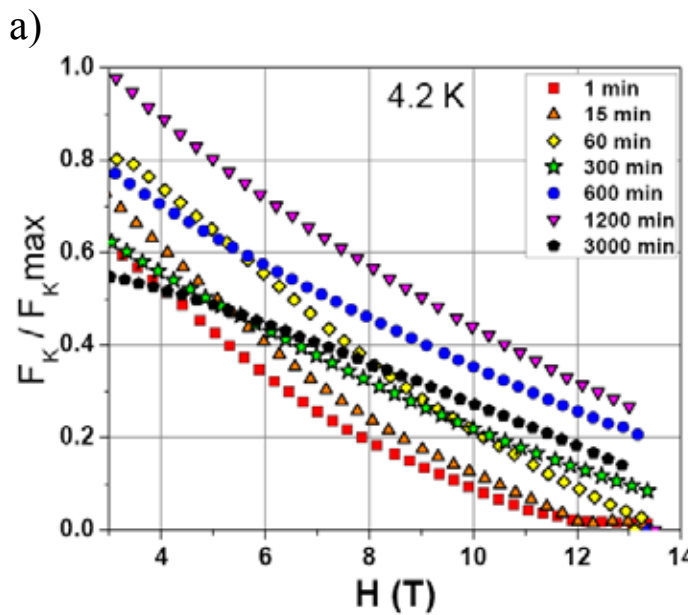

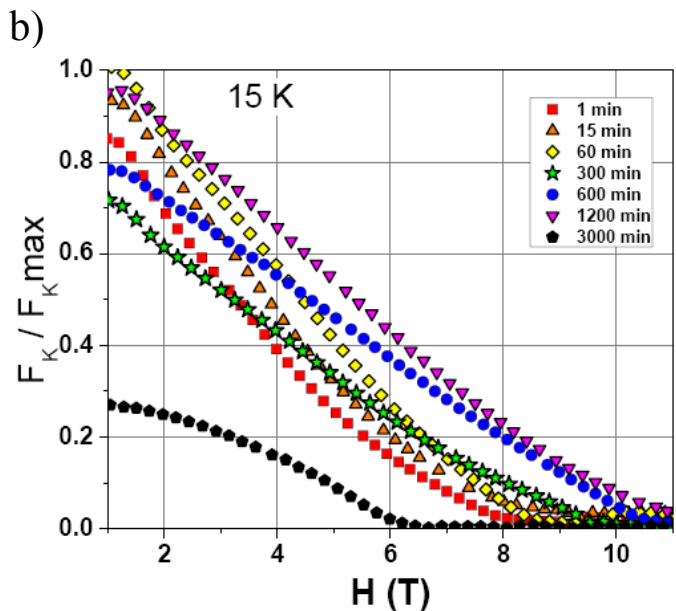

**Figure 7** –Kramer function ($F_K = J_c^{0.5}H^{0.25}$) normalized to the highest value measured, as a function of applied field H at a) 4.2 K and b)15 K. Kramer plots became increasingly linear at longer milling times, permitting determination of H* by extrapolation.

$J_c$ was calculated from M-H loops using equation (1). $J_c$(15 K) and $J_c$(4.2 K) are shown in Figure 5. Due to experimental limitations (mainly flux jumping), the 4.2 K curves do not extend to zero field, but the 15 K $J_c$(H) curves (Fig 5a) show that the sample milled for 60 min had the best low-field $J_c$. Maximum $J_c$(0 T, 15 K) of each sample (in order of ascending milling time) was: 392, 472, 548, 290, 335, 483, and 76 kA/cm$^2$. Although low-field $J_c$ had little relation to milling time, high field (8 T) $J_c$ benefited strongly from longer milling times up to 1200 min, because milling enhanced H* (see below) and thus reduced the dependence of $J_c$ on field. The 1200 min sample had only slightly higher $J_c$(3 T, 4.2 K) than the 60 min sample, but at 12 T the difference was more than an order of magnitude.

Figure 6 shows pinning force curves at 4.2 K and 15 K. As in the case of $J_c$, the curves in Figure 6a show that the pinning force dependence on milling time was not straightforward. We did observe $F_p$ to generally increase with increased milling time, except for the very strong decrease between 60 min and 300 min, for which there is no clear explanation, and the even stronger decrease between 1200 min and 3000 min, which we address in the next section. Maximum pinning force at 4.2 K occurred at $t_{mill}$ = 1200 min, and was ~12 GN/m$^3$. Because of flux jumping, we did not measure accurate $F_p$max values for all samples at 4.2 K. Accordingly, Figure 6b gives $F_p/F_p$max values at 15 K, plotted against reduced field (H/H*). Changes to the shape of the pinning force curve were not entirely systematic, but increased milling time generally resulted in a slight shift of the peak position to higher reduced field, while increasing the normalized pinning force at mid-range H/H* values. The curve for the 3000 min sample had by far the most favorable curve shape, with a peak at H/H* = 0.31, rather than the H/H* ~0.2 measured for the other samples. Furthermore, the 3000 minute sample retained much greater normalized pinning force at high H/H* than any of the other samples. Unfortunately, we have seen that $t_{mill}$ = 3000 min also resulted in reduced $T_c$ and $H_{c2}^{//}$, and we will show later that such a long milling time was also detrimental to connectivity.

Figure 7 shows that Kramer function [37] plots ($F_K$ = $J_c^{0.5}H^{0.25}$) at 4.2 K are nearly linear, particularly for long milling times, which enables the irreversibility field (H*) to be determined by extrapolating the Kramer line to the field axis intercept. As noted later we believe that H*(T) is closely related to the lower value of $H_{c2}$, $H_{c2}^{\perp}$, associated with those grains whose ab-planes lie perpendicular to the applied field direction.

Figure 8 plots H* as a function of milling time at 4.2 K and 15 K defined using two criteria: H at which $J_c$ = 100 A/cm$^2$ and the extrapolated Kramer line intercept [25]. H*(4.2 K) increased from 12 T at $t_{mill}$ = 1 min up to 17.2 T at $t_{mill}$ = 1200 min. Further milling resulted in no further increase in H*(4.2 K), and by $t_{mill}$=3000 min, H*(4.2 K) decreased slightly to 16.2 T. At 15 K, H* peaked around 11 T for $t_{mill}$ = 1200 min and then declined at longer milling time, consistent with $H_{c2}$(T) and $T_c$ data given above, which indicate that the effects of $T_c$ suppression were most pronounced at $t_{mill}$ = 3000 min.

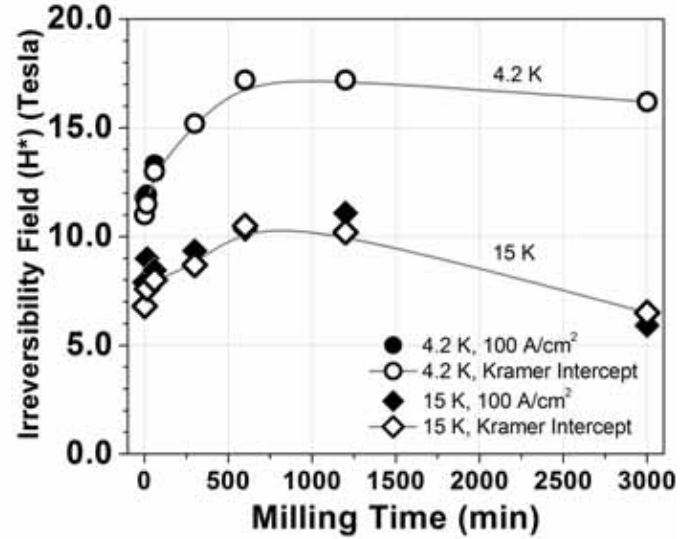

**Figure 8** – Irreversibility field (H*) as a function of milling time at 4.2 K (circles) and 15 K (diamonds) using $J_c$ = 100 A/cm$^2$ criterion (solid symbols) or Kramer line intercept (hollow symbols). Lines are a guide to the eye.

**Discussion**

The goals of this work were 1) to determine the duration of milling necessary to alloy the material with C and 2) to investigate the nature and progression of non-compositional superconducting property enhancements produced by heavy milling. In order to isolate the effect of progressive milling from the effect of C-doping, the bulk composition was fixed at Mg(B$_{0.95}$C$_{0.05}$)$_2$, with fixed heat treatment ($T_{HIP}$=1000$^o$C, $t_{HIP}$ = 200 min). The a-axis lattice parameter was 3.068-3.069 Å for $t_{mill} \geq$ 60 min. In this section, we show that this parameter corresponds to a nearly constant lattice C-content of X~0.05. Since the lattice C-content ceased to vary for $t_{mill}$ > 60 min, the continued improvement of H* and $J_c$ with further milling, we conclude that longer-time changes in properties had non-compositional causes. We found that the low-field $J_c$ did not change systematically with $t_{mill}$, but that high-field $J_c$ enhancements correlated strongly to enhanced H*. Finally, we estimate grain size from XRD peak breadth and correlate H* to inverse grain size, concluding that the observed ~50% H* enhancement is largely due to grain refinement resulting from longer milling times.

Lattice C-content (X) has been shown [19,20,29] to produce a linearly varying a-axis lattice parameter. By comparison to C-doped single crystals [19], we can estimate X for our heat treated samples. Those data are shown in Figure 9. Even 1 min of milling provided enough mixing to achieve a lattice parameter shift corresponding to C-content X ~ 0.03, which increased after 60 min of milling to X ~ 0.05. This was consistent with the X = 0.04 intentionally added and our unpublished observations that the as-received $MgB_2$ powder contained ~ 0.01C. Longer milling times did not result in significant further lattice parameter shifts, suggesting constant C-content of the lattice, in spite of incorporating of WC into the sample as seen in Figure 2.

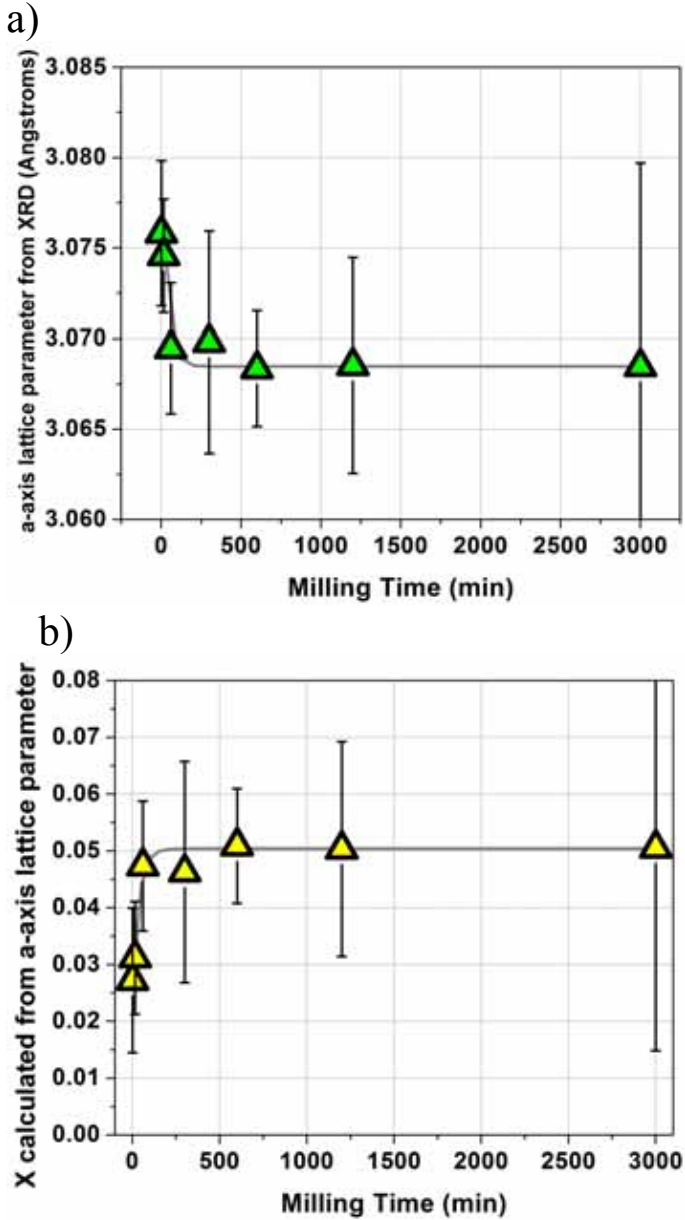

**Figure 9** – a) a-axis lattice parameter (in Å) calculated from XRD peak positions and b) lattice C-content X in $Mg(B_{1-X}C_X)_2$ calculated from the data in Table I.

WC and Co from the milling media entered the powder mixture during milling. No published reports exist of W or Co detected within the $MgB_2$ lattice, nor did our analytical TEM examination detect W or Co within $MgB_2$ grains, so we conclude that those elements have not doped the $MgB_2$ lattice to any significant level. We conclude that changes in properties for samples milled 60 min and longer were therefore due to microstructural factors such as connectivity and grain size.

Connectivity can be assessed by the Rowell analysis [30,12] if one assumes a universal $\Delta\rho_{ideal}$ = $\rho(300)-\rho(residual)$ and then scales the measured resistivity curves assuming that only the connected fraction of the total cross sectional area ($A_F$) carries electrical current. $\Delta\rho_{ideal}$ as a function of carbon content can be taken from the well-connected, randomly-oriented polycrystalline CVD filaments of Wilke et al. [16,17] as discussed elsewhere.[12] When we match our calculated lattice C-content to the CVD filaments, and apply $\Delta\rho_{ideal}$ to the data in Table II (using C-contents from Figure 9b), we obtain the data in Table III.

**Table III** – Rowell analysis results using $\Delta\rho(\mu\Omega\text{-cm})$ = 109.59**X** + 8.8794 where **X** is the mole fraction of C in $Mg(B_{1-X}C_X)_2$.[12,16,17]

| Milling Time (minutes) | X From XRD | $\Delta\rho_{ideal}$ from CVD filaments ($\mu\Omega$-cm) | $A_F$ | $\rho_A$ (40 K) ($\mu\Omega$-cm) |
|---|---|---|---|---|
| 1 | 0.027 | 11.9 | 0.62 | 9 |
| 15 | 0.031 | 12.3 | 0.68 | 11 |
| 60 | 0.047 | 14.1 | 0.61 | 20 |
| 300 | 0.046 | 14.0 | 0.56 | 32 |
| 600 | 0.051 | 14.5 | 0.58 | 42 |
| 1200 | 0.050 | 14.4 | 0.52 | 54 |
| 3000 | 0.050 | 14.4 | 0.35 | 99 |

Presuming 10 vol% porosity and 10 vol% non-superconducting second phases, we would expect a maximum $A_F$ = 0.8. Here we see that the $A_F$ started at ~0.65 for short $t_{mill}$, then decreased slowly as $t_{mill}$ increased, falling to only 0.35 at $t_{mill}$ = 3000 min. The microstructural reason for the decrease is not clear. Only the sample with $t_{mill}$ = 3000 min had $A_F$ < 0.5. This probably accounts for the 3000 min sample having the lowest low-field $J_c$ of the set. However, in the milling time regime (60 min ≤ $t_{mill}$ ≤ 1200 min, which defines the region of non-compositional $J_c$(8 T, 4.2 K) increase with increasing milling time) $A_F$ decreased by only 0.09, and was not the main factor determining $J_c$, even at low field. In fact, low-field (H < 1 T) $J_c$(15 K) bore no apparent relation to $A_F$, except in the 3000 min sample that had both low $A_F$ and low $J_c$.

The $J_c$(H,T) behavior of these samples was close to Kramer-like (straight lines in Figure 7), suggesting the applicability of the flux line lattice shear model (FLL)[31,32,37] following the general form:

$$(2) \quad F_P \propto b^{\frac{1}{2}}(1-b)^2$$

where $b = B / B^*$. Since $F_P = J_c B$, and $B = \mu_o H$, and $MgB_2$ is assumed to follow the same $J_c \propto 1/d$ (where d is grain size) behavior as $Nb_3Sn$ [31], and incorporating $J_c \propto A_F$ (the active cross sectional fraction), we obtain the following relationship to describe $J_c$ as a function of applied field, grain size, connectivity, and irreversibility field.

$$(3) \quad J_C \approx \frac{C A_F}{d} h^{-\frac{1}{2}}(1-h)^2$$

Where $C$ is a constant, $A_F$ is the active cross section, $d$ is grain size, and $h = H / H^*$. [31,32] The constant $C$, $A_F$, and grain size are assumed to be field-independent. As H approaches H*, $J_c$ is strongly dependent on H*. For example, imagine Sample A with H* = 13.5 T and Sample B with H* = 17.2 T. At 8 T, $J_c(A) \sim 0.22C$, and $J_c(B) \sim 0.42C$ – almost twice as high. At 10 T, the difference would be a factor of about 3. Thus differences in $J_c$ at high fields are more clearly linked to variations of H* than to variations of connectivity. Therefore, in most sample sets, the single largest factor determining mid to high-field $J_c$ is H*. We will argue below that H* is dominated by grain size.

XRD peak widths (FWHM) of sintered samples (including data from Table I) were subjected to Williamson-Hall analysis to determine grain size and microstrain. It should be remembered that instrument broadening can also contribute to FWHM. Therefore grain thickness from this analysis should be viewed as a *minimum* estimate of actual grain size. The calculated grain sizes and microstrains are given in Figure 10 as a function of $t_{mill}$. Calculated grain size decreased rapidly to 23 nm for $t_{mill} \leq 600$ min and then more slowly (if at all) thereafter. Microstrain was only 0.0013 - 0.0021 and generally decreased with increased milling time.

These grain size data agreed fairly well with the TEM results (Figure 2) which found grain size <30 nm in samples with $t_{mill}$=1200 and 3000 min. Grain boundaries are important for the superconducting properties because they both pin flux vortices and raise $H_{c2}$ by scattering electrons.[33] Grain boundary density (boundary area per unit volume) goes as the inverse of grain size (1/d).

Figure 11 shows that the measured H*(4.2 K) scales linearly with 1/d, and that a nearly 50% increase in H*(4.2 K) from 11.8 to ~17.2 T was obtained by reducing grain size about 75%, due either to enhanced electron scattering at grain boundaries or enhanced flux pinning.[36] In order to determine which mechanism is more likely, we can consider two additional pieces of evidence. 1) Improved flux pinning would be expected to improve both H* *and* low-field $J_c$. But in fact there was no clear relationship between $t_{mill}$ and $J_c$(0 T, 15 K) or $F_{pmax}$ (4.2K) as Figs. 5 and 6 show. 2) By contrast, the electron-scattering-sensitive properties $T_c$, $dH_{c2}/dT$, and normal state resistivity changed strongly as a function of $t_{mill}$.

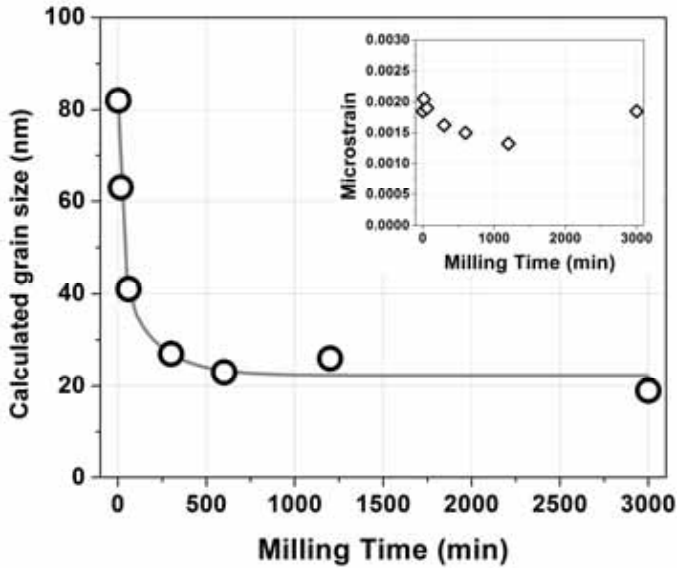

**Figure 10** – Grain size decreased rapidly as a function of increased milling time, as calculated by Williamson-Hall analysis of XRD peak widths. Gray line is a guide to the eye. Inset plot shows microstrain as a function of milling time, as calculated by the same analysis.

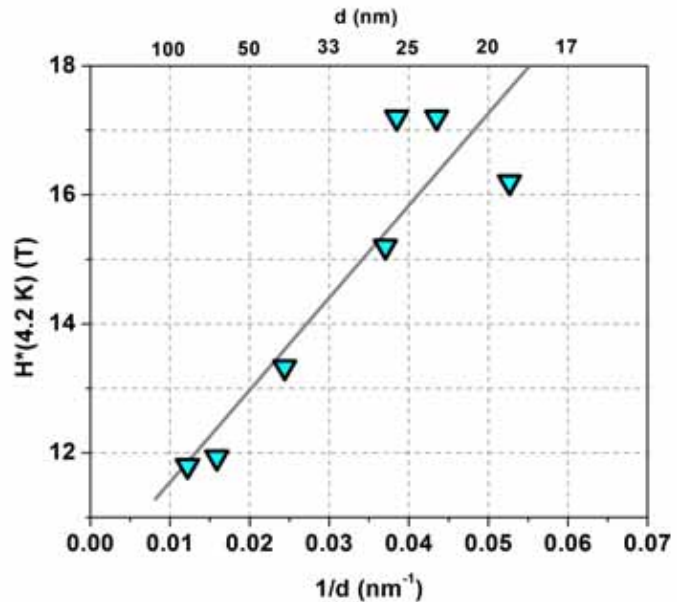

**Figure 11** – Linear relationship between H* (4.2 K) from Figure 8 and inverse grain size.

Typically, H* is a substantial fraction (perhaps 80%) of the pertinent $H_{c2}$ value. Figure 4 shows the perhaps surprising result that $H_{c2}^{//}$ was essentially unchanged at milling times between 60 min and 1200 min, despite the strong increase in H* over the same milling time range. Because H* changes independently of $H_{c2}^{//}$ we conclude that increased H* results from increased $H_{c2}^{\perp}$. Approximating $H_{c2}^{\perp}$ by H*, the ratio of the measured $H_{c2}$ to H* is approximately the $H_{c2}$ anisotropy $\gamma$, which can be >5 for pure $MgB_2$.[28] Here we find that $\gamma$(4.2 K) is then approximately 33/13 ~ 2.5 for 60 min, 33/15 ~ 2.2 for 300 min, 33/17 ~ 2 for 1200 min and 25/16 ~ 1.5 for the 3000 minute sample. The important result is that *grain refinement* by milling for 60 min $\leq t_{mill} \leq$ 1200 min *improved $H_{c2}^{\perp}$* (and therefore H* and $J_c$(H)) *without any significant change to $H_{c2}^{//}$* compared to single crystals and unmilled bulks. This improvement is exactly that desired for bulk applications of $MgB_2$ since the effective domain of critical current performance in untextured samples is always constrained by the full critical state irreversibility field determined by the lower of the two critical fields, that is $H_{c2}^{\perp}$

The observation that long $t_{mill}$ can be responsible for very fine <30 nm grains even after sintering at 1000°C requires explanation. In principle, one expects that the driving force for grain growth is increased by both grain refinement and cold work, leading to an expectation that grain growth should be rapid when HIPing fine-grained, heavily milled crystals. There are two possibilities why the grains remained fine: either grain boundary mobility was limited by slow diffusion kinetics or grain boundaries were pinned by second phases. Concerning the possibility that 200 min at 1000°C was insufficient for diffusional grain growth, we note that 1000°C is indeed < 0.5 $T_m$ but also note that Wilke et al. [34,35] showed clear evidence of point defect annealing at T < 700°C, which would lead to the expectation of reasonable diffusion rates at 1000°C. We thus believe that significant barriers to grain boundary growth are present.

Observations by TEM of some extremely large grains (1 μm compared to 20-30 nm in matrix) in the most heavily milled sample (Fig 2b) appear to constitute an example of abnormal (discontinuous) grain growth – a phenomenon known to occur when normal grain growth is partly constrained by an inhomogeneous distribution of fine precipitates. Nano-porosity, WC rubble, and MgO suitable for such inhibition of grain growth were all observed by TEM. The finely dispersed MgO may have resulted from atmospheric reaction layers originally present on powder particle surfaces, pulverized by milling. The anomalously large grains occasionally seen in the TEM then occur in regions where such particles are sparsely distributed and grain growth does, anomalously, become possible.

## Conclusions

In this work we determined the effect of milling time on alloying, connectivity, and grain size of $MgB_{1.95}C_{0.05}$ in order to optimize $J_c$(H,T). We found that only 60 min of milling was necessary to incorporate carbon dopant up to the lattice composition X = 0.05, but that grain size continued to decrease to the very small value of 20-30 nm with additional milling time. As a result of this grain refinement, H*(4.2 K) increased strongly with increasing $t_{mill}$ from 13.3 T (60 min) to 17.2 T (1200 min), probably due to grain boundary scattering improving $H_{c2}^{\perp}$. This H* increase was the primary non-compositional benefit of milling, and resulted in doubling of $J_c$(8 T, 4.2 K) in samples with similar composition. By assessing the connectivity using the Rowell analysis, we found that $J_c$(8 T, 4.2 K) was much more strongly influenced by H* than by connectivity ($A_F$), which was only a weak function of milling time. Therefore we conclude that grain refinement was the primary mechanism by which high energy milling increased $J_c$(8 T, 4.2 K) to a peak value of ~ 8.5x10$^4$ A/cm$^2$ for $t_{mill}$=1200 min. The perhaps surprising result that 20-30 nm grain sizes could be maintained through the 1000°C heat treatment is explained by the inhibition of grain growth by finely dispersed second phases produced as a side-effect of the milling process.

## Acknowledgements

This work was performed under a Fusion Energy Sciences Fellowship administered by DOE through Oak Ridge Institute for Science Education with additional support from the DOE – Office of High Energy Physics and the NSF Focused Research Group on Magnesium Diboride (FRG): two-gap superconductivity in magnesium diboride and its implications for applications, DMR-0514592. We are grateful to Alex Squitieri, Van Griffin, William Starch, Jan Jaroszynski who helped with high field measurement, and R. Wilke of Iowa State University who provided us with additional resistivity data on CVD-made filaments. Additional thanks to Durval Rodrigues Jr., A.